# I-SolFramework: An Integrated Solution Framework Six Layers Assessment on Multimedia Information Security Architecture Policy Compliance

Heru Susanto[123], Mohammad Nabil Almunawar[1], Yong Chee Tuan[1] and Mehmet Sabih Aksoy

[1]FBEPS University of Brunei
Information System Security Group
susanto.net@gmail.com

[2]The Indonesian Institute of Sciences
Information Security & IT Governance Research Group
heru.susanto@lipi.go.id

[3]King Saud University
Department of Information System
hsusanto@ksu.edu.sa

***Abstract.*** – Multimedia Information security becomes a important part for the organization's intangible assets. Level of confidence and stakeholder trusted are performance indicator as successes organization, it is imperative for organizations to use Information Security Management System (ISMS) to effectively manage their multimedia information assets. The main objective of this paper is to Provide a novel practical framework approach to the development of ISMS, Called by the I-SolFramework, implemented in multimedia information security architecture (MISA), it divides a problem into six object domains or six layers, namely organization, stakeholders, tool & technology, policy, knowledge, and culture. In addition, this framework also introduced novelty algorithm and mathematic models as measurement and assessment tools of MISA parameters.

## I. INTRODUCTION

The development of IT security framework has gained much needed momentum in recent years, there continues to be a need for more writings on best theoretical and practical approaches to security framework development (*potter & beard, 2010*). Thus, securing information system resources is extremely important to ensure that the resources are well protected. Information security is not just a simple matter of having usernames and passwords (*solms & solms, 2004*). This paper proposed novel framework as information security assessment, might be implemented in various kinds, types and sizes of object or organization, called by integrated solution framework (I-SolFramework), which has various characteristics, including six major domains or layers, namely: Organization, Stakeholders, Tools & Technology, Policy, Culture and Knowledge. At the end of the paper, I-SolFramework would implement on multimedia information security architecture (*susanto & muhaya, 2010*) to assessing and measuring organization readiness level for multimedia information security compliance.

## II. RELATED WORK

In this chapter we discussed several framework issues as state of the art as author's development guidance to introduced new paradigm of information security framework. Information Security Governance Framework (*ohki, 2009*) proposed Information Security Governance Framework which combines and inter-relates existing information security schemes. (*eloff & eloff, 2005*) Introduced security architecture includes process of developing risk awareness, the assessment of current controls, and finally, alignment of current and new controls to meet the organization's information security requirements. Integrated information security architecture (ISA) is the mechanism to ensure that all individuals know their responsibilities and how they need to go about protecting the company's information security resources. Multimedia Information Security Architecture (*susanto & muhaya, 2010*) introduced us eight layers as MISA as main parameters of multimedia security [*figure 1*].

### III. MULTIMEDIA INFORMATION SECURITY ARCHITECTURE

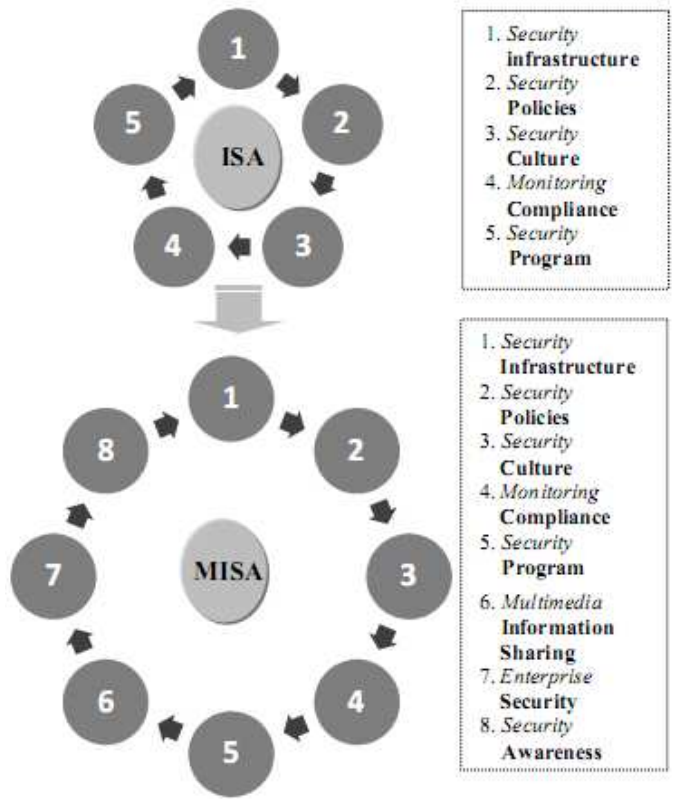

*Figure 1.* Multimedia Information Security Architecture –MISA

The framework is emphasizing on information security in multimedia issue, into existing architecture. Based on ISA architecture, it has five main components, authors' proposed MISA architecture which has eight major components, called by MISA controls, namely;

***Security Infrastructure*** indicated all components of the infrastructure in organization affected to the information security, especially those directly related to multimedia information.

***Security Policies*** defined as measures taken by an organization with respect to their information security. Some of the information security standard provides a comprehensive overview of security policies, the top five information security standards that exist and widely used are BS 7799, ISO 27001, COBIT, ITIL, PCIDSS (*susanto, almnawar & tuan, 2011b*).

| Control Title | Structure of MISA ||
|---|---|---|
| | Control Number | Section |
| Security Program | 5 | 5.1. |
| | | 5.2. |
| | | 5.n. |
| Security Awareness | 8 | 8.1. |
| | | 8.2. |
| | | 8.n. |
| Security Infrastructure | 1 | 1.1. |
| | | 1.2. |
| | | 1.n. |
| Enterprise Security | 7 | 7.1. |
| | | 7.2. |
| | | 7.n. |
| Multimedia Information Sharing | 6 | 6.1. |
| | | 6.2. |
| | | 6.n. |
| Security Policies | 2 | 2.1. |
| | | 2.2. |
| | | 2.n. |
| Security Culture | 3 | 3.1. |
| | | 3.2. |
| | | 3.n. |
| Monitoring Compliance | 4 | 4.1. |
| | | 4.2. |
| | | 4.n. |

*Table 1.* Multimedia Information Security Architecture –MISA

***Security Culture*** *is* recognizing the importance of bringing collaboration and governance process. The values and behaviors that contribute to the unique social and psychological environment of an organization, its culture is the sum total of an organization's past and current assumptions (*BDO*).

***Monitoring Compliance*** is reconciling exist multimedia information with the benchmarking of standard. Standard is expected to guide and limited on the object and domain.

***Security Program*** stated to make recommendations for improving the security of computer systems and the information residing on them and provide security initiative recommendations and priorities, and to perform high level threat and risk analysis.

*Multimedia Information Sharing* indicated critical relationships among key of multimedia information resource to share information with other important and relevant components.

*Enterprise Security* the practice of applying a comprehensive and rigorous method for describing a current and/or future structure and behavior for an organization's security processes, information security systems, personnel and organizational sub-units, so that they align with the organization's core goals and strategic direction (*gardner*).

*Security awareness* is the knowledge and attitude members of an organization possess regarding the protection of the physical and, especially, information assets of that organization (*gardner*).

## IV. SIX LAYERS ASSESSMENT FRAMEWORK: I-SOLFRAMEWORK

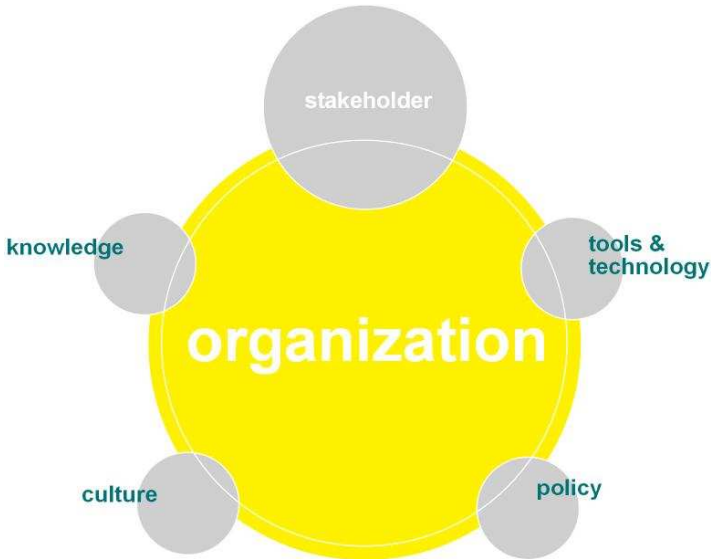

*Figure 2.* Proposed framework

This section we introduced new framework for approaching object and organization analyst, called by I-SolFramework, abbreviation from *I*ntegrated *Sol*ution for Information Security

*Framework*. The framework consists of six layers component *[figure 2]*: organization, stakeholder, tools & technology, policy, culture, knowledge. We introduced the basic elements of development in I-SolFramework profile as illustrated. These elements are identified in the following subsection (elucidation of terms and concepts)

## V. ELUCIDATION OF TERM AND CONCEPT

- **Organization:** A social unit of people, systematically structured and managed to meet a need or to pursue collective goals on a continuing basis, the organizations associated with or related to, the industry or the service concerned (*BDO*).

- **Stakeholder:** A person, group, or organization that has direct or indirect stake in an organization because it can affect or be affected by the organization'sactions, objectives, and policies (*BDO*).

- **Tools & Technology:** the technology upon which the industry or the service concered is based. The purposeful application of information in the design, production, and utilization of goods and services, and in the organization of human activities, divided into two categories (1) Tangible: blueprints, models, operating manuals, prototypes. (2) Intangible: consultancy, problem-solving, and training methods (*BDO*).

- **Policy:** typically described as a principle or rule to guide decisions and achieve rational outcome(s), the policy of the country with regards to the future development of the industry or the service concerned (*BDO*).

- **Culture:** determines what is acceptable or unacceptable, important or unimportant, right or wrong, workable or unworkable. *Organization Culture:* The values and behaviors that contribute to the unique social and psychological environment of an organization, its culture is the sumtotal of an organization's past and current assumptions (*BDO*).

- **Knowledge:** in an organizational context, knowledge is the sum of what is known and resides in the intelligence and the competence of people. In recent years, knowledge has come to be recognized as a factor of production (*BDO*).

## VI. ALGORITHM AND MATHEMATICS MODEL FOR ASSESSING PARAMETERS OF MISA

Assessment is a deterministic process, which is ubiquitously present in the world. Assessment tools on measuring MISA controls on organization readiness level, MISA control and structure content is mapping to the six layers framework are provided *[figure 3]*.

| Six Layer on Framework | Structure of MISA | |
|---|---|---|
| | **Title** | **Control Number** |
| **Organization** | Security Program | 5 |
| **Stakeholder** | Security Awareness | 8 |
| **Tool & Technology** | Security Infrastructure | 1 |
| | Enterprise Security | 7 |
| **Policy** | Multimedia Information Sharing | 6 |
| | Security Policies | 2 |
| **Culture** | Security Culture | 3 |
| **Knowledge** | Monitoring Compliance | 4 |

*Table 2.* Multimedia Information Security Architecture –MISA

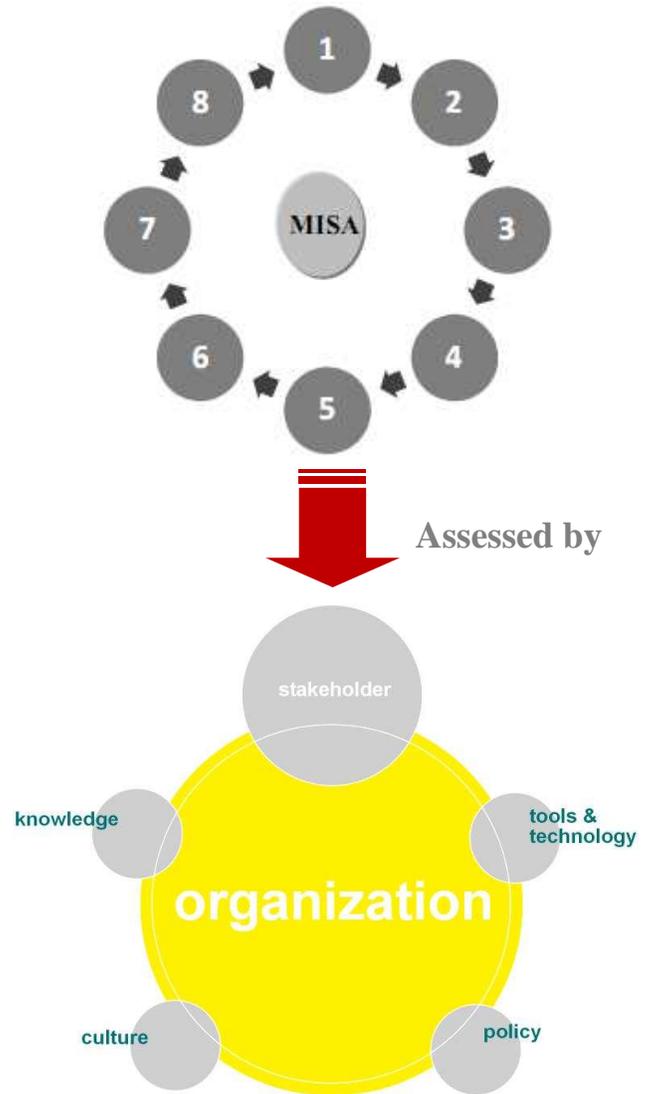

*Figure 3.* Proposed framework

### VI.I. ALGORITHM

Six layer framework algorithms is calculating eight controls of MISA, as parameter, trough framework layers, measurement results is an indicator for an organization readiness for MISA compliance. Algorithm works based on lowest level calculation, bottom up approach of estimation, of sub control *[figure 4]*. Details of 1$^{st}$ level and 2$^{nd}$ level of algorithm are presented below.

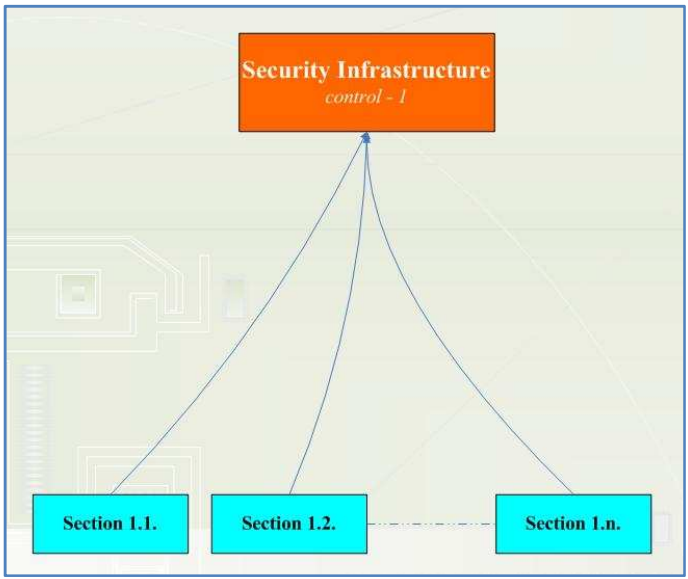

*# Algorithm level- 2 for measuring Section issues #*

*Initiate;*
*Step 1 [measuring section]*
*SectionIssue#1, sectionIssue #2,..., sectionIssue#n*
            *as property of dependent Control;*
*TotalSectionIssue =*
*SectionIssue #1 + SectionIssue #2 + ... + SectionIssue #n;*
*ValControl#1 = TotalSectionIssue / n;*
*Control#1 = ValControl#1*
*# end #*

*# Algorithm level-1 for measuring assessment control #*
*Initiate;*
*Step 2 [measuring control]*
*Coltrol#1, Control#2,...,Control#n as property of dependent MISA;*
*TotalValControl = ValControl#1 + ValControl#2 + ... + ValControl#n;*
*ValMISA = TotalValControl / n;*
*MISA = ValMISA*
*# end #*

*Figure 4.* Algorithm 2$^{nd}$ level for measuring assessment issues

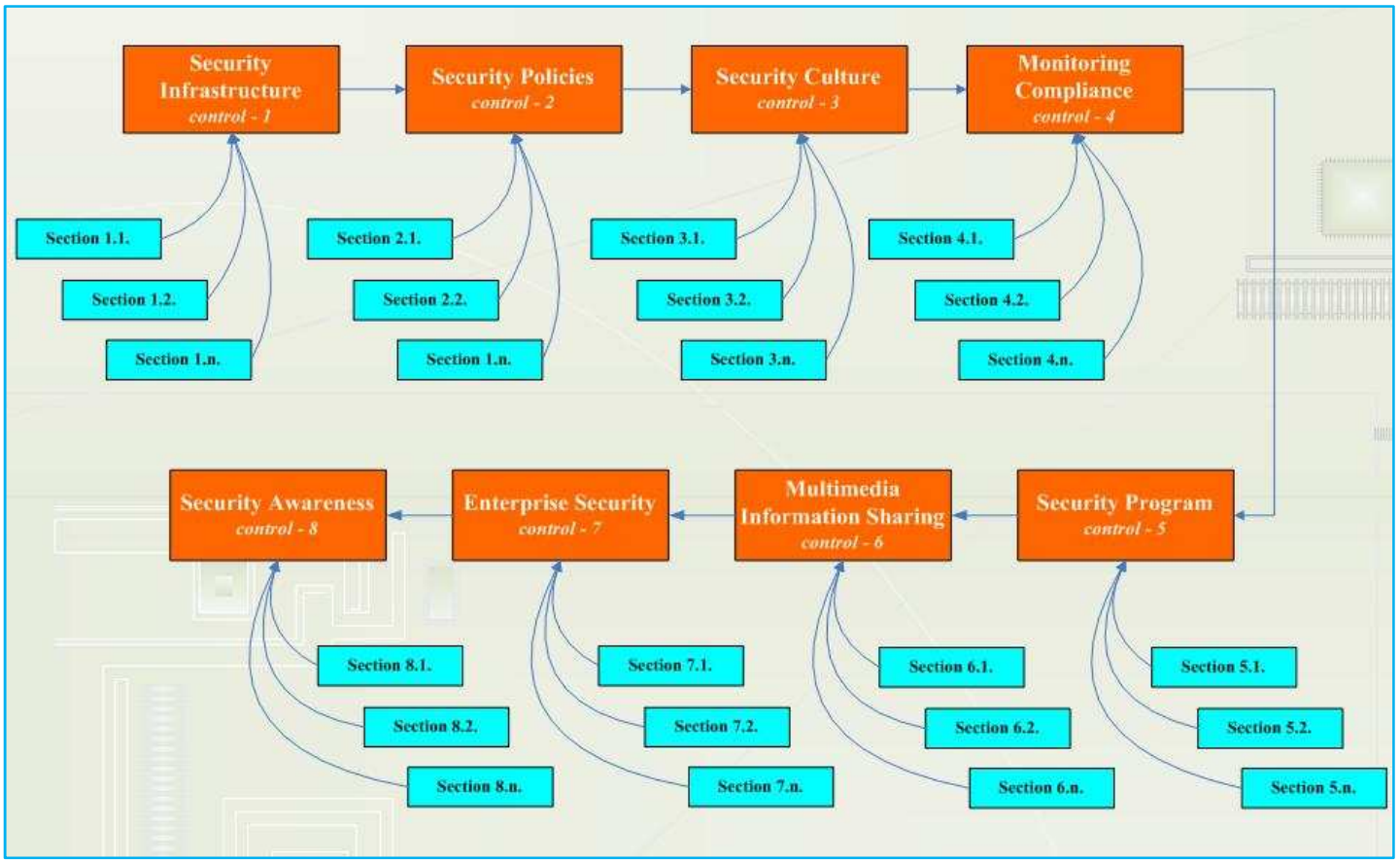

*Figure 5.* Algorithm 1$^{st}$ level for measuring assessment issues

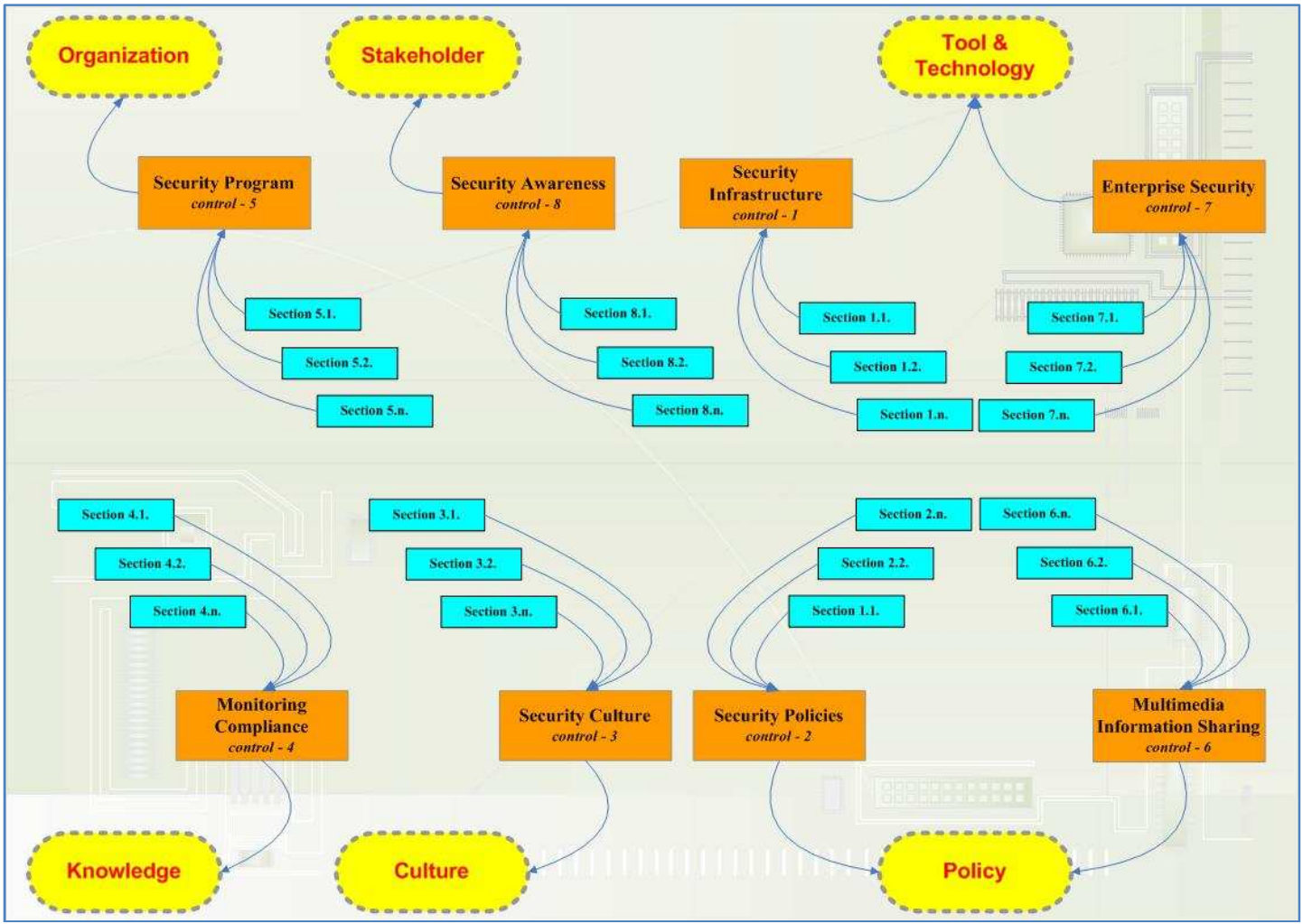

*Figure 6.* MISA controls on six layers framework

*Figure 6*, explain us how controls of MISA is mapped to six layers framework, I-SolFramework, table 2 descript each control, section and top layer domain of framework is given. Table 2 summarizes mapping results of all layers together with their associated controls (*alfantookh, 2009*), (*susanto et al, 2011a*) and (*susanto, almunawar and tuan, 2011b*) based on six layers framework.

## VI.II. MATHEMATICAL MODEL

The mathematical model is explaining us as well as facilitating readers gain a comprehensive and systematic overview of the mathematical point of view. Modeling begins by calculated lowest level components, namely *section*. Determining the lowest level could be flexible, depending on the problems facing the object, might be up to $3^{rd}$, $4^{th}$, $5^{th}$ ... $N^{th}$ level, for MISA case study, authors define problem until $2^{nd}$ level.

Formula works recursively; enumerate value from the lowest level, until the highest level of framework. several variables used as position and contents of framework indicator, where *k* as *control*, *j* as *section*, and *h* as a *top level*, details of these models are mention as follows.

$$(a) \rightarrow x_j = \sum_{k=1}^{n} \frac{[section]_k}{n}$$

$x_j: control$

**(a) → $x_j$** Indicate value of control of MISA which is resulting from *sigma* of section(s) assessment, divided by number of section(s) contained on the lowest level.

$$(b) \rightarrow x_i = \sum_{j=1}^{n} \frac{[control]_j}{n}$$

$x_i: domain$

**(b) → $x_i$** Stated value of domain of MISA which is resulting from *sigma* of control(s) assessment, divided by number of control(s) contained at concerned level.

After *(a), (b),* defined, next step is substituted of mathematical equations, into new comprehensive modeling notation in a single mathematical equation, as follows.

$$x_h = \sum_{i=1}^{n} \frac{[control]_i}{n}$$

$$x_h = \sum_{i=1}^{n} \frac{[b]_i}{n}$$

$$x_h = \sum_{i=1}^{n} \frac{\left[\sum_{j=1}^{n} \frac{[section]_j}{n}\right]_i}{n}$$

So that for six layer, or we called it by top level, equation will be:

$$x_h = \sum_{i=1}^{6} \frac{\left[\sum_{j=1}^{n} \frac{\left[\sum_{k=1}^{n} \frac{[control]_k}{n}\right]_j}{n}\right]_i}{6}$$

*Where; k=section; J=control; I=domain (organization, stakeholder, tools & technology, policy, knowledge, and culture).*

The algorithm is considered to be reliable and easy implementing in analyzing such problem, emphasized on divided problems into six layers as the initial reference in measuring and analyzing the object (*susanto et al, 2011c*) and (*susanto, almunawar and tuan, 2012*). The results obtained as full figure indicators of an organization's readiness in multimedia security compliance, it showed us strong and weak point on layer of the object. Indicated, layer with a weak indicator has a high priority for higher visibility in the framework of improvement and perfection of the system as a whole. In the manner of the six layer framework, I-SolFramework, analysis could be done easily and simply observe.

## VII. AN ILLUSTRATIVE MEASUREMENT

An illustrative example has been considered in order to illustrate used of approach. Each control, section and top layer domain of framework, a value associated with the example is given. Table 3 summarizes the results of all layers together with their associated controls based.

| Six Layer on Framework | Structure of MISA | | |
|---|---|---|---|
| | Title | Control Number | Assessment Result |
| Organization | Security Program | 5 | 54.5 |
| Stakeholder | Security Awareness | 8 | 50 |
| Tool & Technology | Security Infrastructure | 1 | 51.1 / 53.5 |
| | Enterprise Security | 7 | 55.8 |
| Policy | Multimedia Information Sharing | 6 | 85 / 78.5 |
| | Security Policies | 2 | 72 |
| Culture | Security Culture | 3 | 47.5 |
| Knowledge | Monitoring Compliance | 4 | 59 |
| Overall Score | | | 57.2 |

*Table 3.* Multimedia Information Security Architecture –MISA

The results given in table 3, are illustrated in the following figures. Figure 1 illustrates the state of the six layers based on MISA controls assessment result. The overall score of all domains is shown in the table to be "57.2 points". The layer of "policy" scored highest at "78.5", and the domain of the "culture" scored lowest at "47.5". Ideal and priority figures are given to illustrate the strongest and weaknesses in the application of each control *[figure 7]*.

*"Ideal"* stated the highest value that can be achieved by an organization, based on six layers framework mentioned.

*"Achievement"* declared the performance of an organization which is the result of the measurement by the framework.

*"Priority"* indicated the gap between ideal values with achievement value. "Priority" and "achievement" it showed inverse relationship. If achievement is high, then domain has a low priority for further work, and conversely, if achievement is low, then the priority will be high.

Priority status or it could be referred as scale of priorities, is the best reference for evaluating of the object or organization based on six main domains. Scale of priority value helping us on evaluating, auditing and maintaining a problem, becomes easier and highly precise.

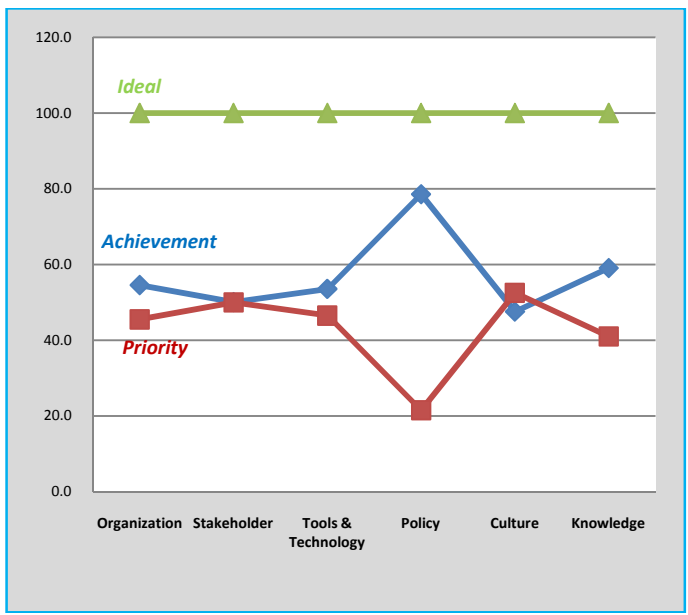

*Figure 7.* Algorithm 1$^{st}$ level for measuring assessment issues

## VIII. CONCLUSION REMARKS

We discussed and showed definitions, step by step how we came ISMS and MISA novelty approach. The approach is about how to finding and collaborate ISMS framework, namely six layers assessment framework, I-SolFramework, to assessing multimedia security controls, which is contains of eight main controls and sub control / sections, works presented in this paper has emphasized the application of new algorithm and mathematic model in measuring readiness level on that multimedia security controls and sections.

## IX. ACKNOWLEDGMENT

This research is supported by FBEPS University of Brunei Darussalam, The Indonesian Institute of Sciences and Information System Department – CCIS, King Saud University.

## X. REFERENCES

Abdulkader Alfantookh. *An Approach for the Assessment of The Application of ISO 27001 Essential Information Security Controls*. Computer Sciences, King Saud University. 2009.

Basie von Solms & Rossouw von Solms. 2004. *The 10 deadly sins of Information Security Management*. Computer & Security 23(2004) 371-376. Elsevier Science Ltd.

Business Dictionary online (BDO). Obtained from: businessdictionary.com

Chris Potter & Andrew Beard. *Information Security Breaches Survey 2010*. Price Water House Coopers. Earl's Court, London.2010.

Eijiroh Ohki, Yonosuke Harada, Shuji Kawaguchi, Tetsuo Shiozaki, TetsuyukiKagaua. *Information Security Governance Framework*. JIPDEC ISMS. 2009

Garnerd, Inc. Wikipedia on Enterprise information security architecture (EISA) and Security awareness.

Heru Susanto & Fahad bin Muhaya. *Multimedia Information Security Architecture*. @IEEE. 2010.

Heru Susanto, Mohammad Nabil Almunawar, Wahyudin P Syam, Yong Chee Tuan, and Saad Hajj Bakry. *I-SolFramework View on ISO 27001. Information Security Management System: Refinement Integrated Solution's Six Domains*. Asian Transaction on Computer Journal. 2011a.

Heru Susanto, Mohammad Nabil Almunawar & Yong Chee Tuan. *Information Security Management System Standards: A Comparative Study of the Big Five*. International Journal of Engineering and Computer Science. IJENS Publishers. 2011b.

Heru Susanto, Mohammad Nabil Almunawar, Yong Chee Tuan, Mehmet Sabih Aksoy and Wahyudin P Syam. *Integrated Solution Modeling Software: A New Paradigm on Information Security Review and Assessment*. International Journal of Science and Advanced Technology. 2011c.

Heru Susanto, Mohammad Nabil Almunawar & Yong Chee Tuan. *I-SolFramework: as a Tool for Measurement and Refinement of Information Security Management Standard*. On review paper. 2012b.

Heru Susanto, Mohammad Nabil Almunawar & Yong Chee Tuan. Information Security *Challenge* and Breaches*:* Novelty Approach on Measuring ISO 27001 Readiness Level. International Journal of Engineering and Technology. IJET Publications UK. Vol 2, No.1. 2012a.

J.H.P. Eloff, M.M. Eloff. *Information Security Architecture.* Computer Fraud & Security. 2005.

Mikko Siponen & Robert Willison. 2009. *Information security standards: Problems and Solution.* Information & Management 46(2009) 267-270. Elsevier Science Ltd.

Shaun Posthumus, Rossouw von Solms. *A framework for the govenance of information security*. Port Elizabeth Technikon, Department of Information and Technology, Private Bag X6011, Port Elizabeth 6000, South Africa

## AUTHORS

**Heru Susanto** is a researcher at The Indonesian Institute of Sciences, Information Security & IT Governance Research Group, also was working at Prince Muqrin Chair for Information Security Technologies, King Saud University. He received BSc in Computer Science from Bogor Agriculture University, in 1999 and MSc in Computer Science from King Saud University, and nowadays as a PhD Candidate in Information Security System from the University of Brunei. 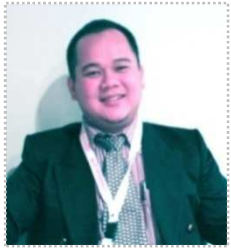

**Yong Chee Tuan** is a senior lecturer at Faculty of Business, Economics and Policy Studies, University of Brunei Darussalam, has more than 20 years of experience in IT, HRD, e-gov, environmental management and project management. He received PhD in Computer Science from University of Leeds, UK, in 1994. He was involved in the drafting of the two APEC SME Business Forums Recommendations held in Brunei and Shanghai. He sat in the E-gov Strategic, Policy and Coordinating Group from 2003-2007. He is the vice-chair of the Asia Oceanic Software Park Alliance. 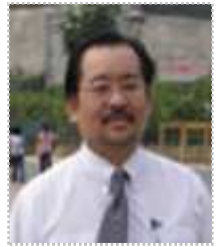

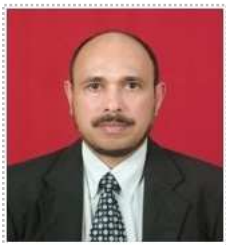**Mohammad Nabil Almunawar** is a senior lecturer at Faculty of Business, Economics and Policy Studies, University of Brunei Darussalam. He received master Degree (MSc Computer Science) from the Department of Computer Science, University of Western Ontario, Canada in 1991 and PhD from the University of New South Wales (School of Computer Science and Engineering, UNSW) in 1997. Dr Nabil has published many papers in refereed journals as well as international conferences. He has many years teaching experiences in the area computer and information systems.

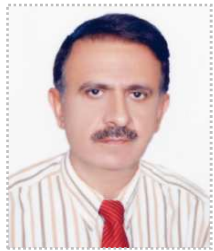**Mehmet Sabih Aksoy** is a Professor in Information System, King Saud Univrsity. He received BSc from Istanbul Technical University 1982. MSc from Yildiz University Institute of Science 1985 and PhD from University of Wales College of Cardiff Electrical, Electronic and Systems Engineering South Wales UK 1994. Prof Aksoy interest on several area such as Machine learning, Expert system, Knowledge Acquisition, Computer Programming, Data structures and algorithms, Data Mining, Artificial Neural Networks, Computer Vision, Robotics, Automated Visual inspection, Project Management.